\documentclass[journal]{IEEEtran}
\usepackage{cite}
\usepackage{amsmath,amssymb,amsfonts}
\usepackage{algorithmic}
\usepackage{graphicx}
\usepackage{textcomp}
\usepackage{caption}
\usepackage{subcaption}
\usepackage{gensymb}
\usepackage{tikz}
\usepackage{pgfplots}

\DeclareMathOperator{\sinc}{sinc}

\pgfplotsset{compat=1.18}

\begin{document}

\title{Cavity Model for a Patch Antenna Embedded with a Hybrid Ground Plane}


\author{\IEEEauthorblockN{Rohan Kalsi,
Angela Nothofer}


\thanks{R. Kalsi and A. Nothofer are with the George Green Institute for Electromagnetics Research, University of Nottingham, Nottingham, NG7 2RD, U.K. (email: eexrk14@nottingham.ac.uk; angela.nothofer@nottingham.ac.uk).
}

}

\IEEEtitleabstractindextext{%
\begin{abstract}
An analytical method to characterise the farfield radiation pattern of a patch antenna embedded with a hybrid ground plane is presented using the Cavity Model. The hybrid ground plane is a two-layered structure with a Perfect Magnetic Conductor (PMC) stacked on top of a Perfect Electric Conductor (PEC) separated by a small distance. The model predicts an increase in the antenna's directivity, which is validated through a full-wave simulation model by realising the PMC surface through an Artificial Magnetic Conductor (AMC) which is embedded into the patch antenna's substrate.
\end{abstract}
}

\maketitle

\IEEEdisplaynontitleabstractindextext

%
\IEEEpeerreviewmaketitle

\section{Introduction}
%
%
%
%
\IEEEPARstart{T}{he} Perfect Magnetic Conductor (PMC) is a structure where the tangential components of the magnetic field vanish next to its surface. An interesting property of an ideal PMC surface is that it exhibits a zero-degree phase shift when an incident electromagnetic waveform is reflected from its surface \cite{AEEBalanis2024}. The concept of a magnetic conductor was initially assumed to not exist in nature and only used in electromagnetic theorems such as the 'Image Theory' and 'Surface Equivalence Theory' to represent virtual sources, and in the 'Induction Theorem' and 'Physical Equivalent Model' to replace physical obstacles in scattering analysis \cite{AEEBalanis2024}. In 1999, Sivenepiper proposed the High Impedance Surface (HIS) ground plane \cite{Sievenpiper1999} which is a combination of two structures, the Artifical Magnetic Conductor (AMC) and the Electromagnetic Bandgap (EBG) limited surface, whereby the AMC has an in-phase reflection characteristic and the EBG surface does not support surface wave propagation. The difference between the HIS and the AMC is the addition of vertical shorting pins which prohibits the propagation of surface waves within the ground plane. The AMC is a realised PMC surface which can be thought of as a surface distribution of scatterers just like that of a Frequency Selective Surface (FSS), but instead of spatial filtering control at a certain frequency band, it is designed for phase reflection control which has applications for enhancing the radiation characteristics of antennas by redirecting back propagating radiation with a controllable phase from -180$^{\circ}$ to 180$^{\circ}$. 
More recently, such surfaces are coined under the term metasurfaces which is described as any periodic two-dimensional structure whose thickness and periodicity are small compared to a wavelength in the surrounding media \cite{Yang2019}. For metasurfaces, resonances occur in the individual scatterers, but not with the periodicity of the array which is what gives EBG structures and FSS their unique properties. Metasurfaces can be also characterised by their effective medium parameters such as their permittivity and permeability, which may exhibit a frequency dependency especially at resonance. 
The AMC can be designed to exhibit destructive interference properties more versatile than a perfect electrical conductor (PEC). An example of such structure is the checkerboard patterned AMC surface \cite{Modi2017b} which has applications in radar cross section (RCS) reduction. This surface redirects the specular scattered waves away from a radar receiver over a large range of frequencies.
Due to properties such as controllable phase reflection and effective medium parameters the AMC has found application in antenna design. Recent applications include ground planes for monopole antennas used in textile on-body antennas to reduce Specific Absorption Rate (SAR) \cite{Alemaryeen2019, Saeed2017, Yan2014}. For patch antennas, the AMC has found applications as a reflector to suppress parallel plate waveguide modes for aperture coupled patch arrays \cite{Zhang2003} and gain enhancement \cite{Yang2013}.

This paper presents the Cavity Model for a rectangular patch antenna embedded with a hybrid ground plane (RPA-HGP), which consists of PMC layer in the middle of the substrate while backed by a perfect electrical conductor (PEC) at its base.
This model provides insight into the radiation characteristics exhibited by the RPA-HGP, which predicts a directivity enhancement when compared to a conventional rectangular patch antenna (RPA). By creating a surface equivalent of the antenna's radiating apertures, the farfield characteristics can be predicted using the Surface Equivalence Theorem. 

Section \ref{surface_equiv_cavity_section} outlines the Surface Equivalence Theorem and Cavity Model, which serves as the foundations before characterising the RPA-HGP.
Section \ref{proposed cavity model} introduces the proposed Cavity Model for the RPA-HGP, detailing the derivation of the electric field components. Section \ref{designing AMC array} presents a realised PMC surface through the miniaturised square ring AMC cells. A circuit equivalent model is created to model the phase reflection properties and this is compared with a full-wave simulation model. Section \ref{amc_integration} realises the hybrid ground plane by embedding the AMC cells into the substrate of the rectangular patch antenna (RPA-AMC) through a full-wave simulation model whereby the farfield patterns are compared to the Cavity Model.

\section{Surface Equivalence Theorem and Cavity Model for a Patch Antenna} \label{surface_equiv_cavity_section}

\begin{figure} [!hb]
    \centering
    \subfloat[\centering ]{{\includegraphics[width=7cm, trim={0cm 0.5cm 0cm 0cm},clip]{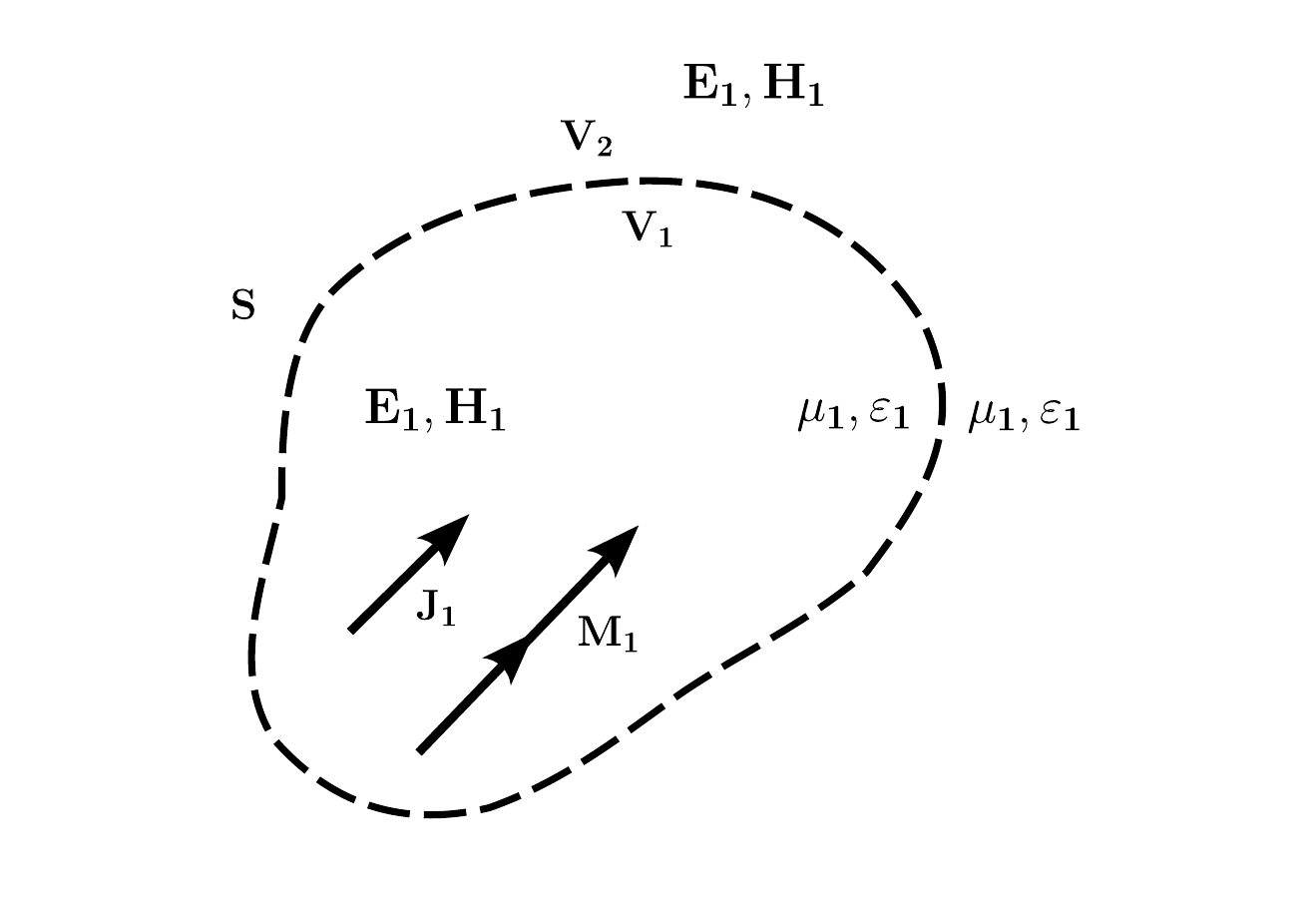} }}%
    \vspace{0.5cm}
    \subfloat[\centering ]{{\includegraphics[width=7cm, trim={0cm 0cm 0cm 0cm},clip]{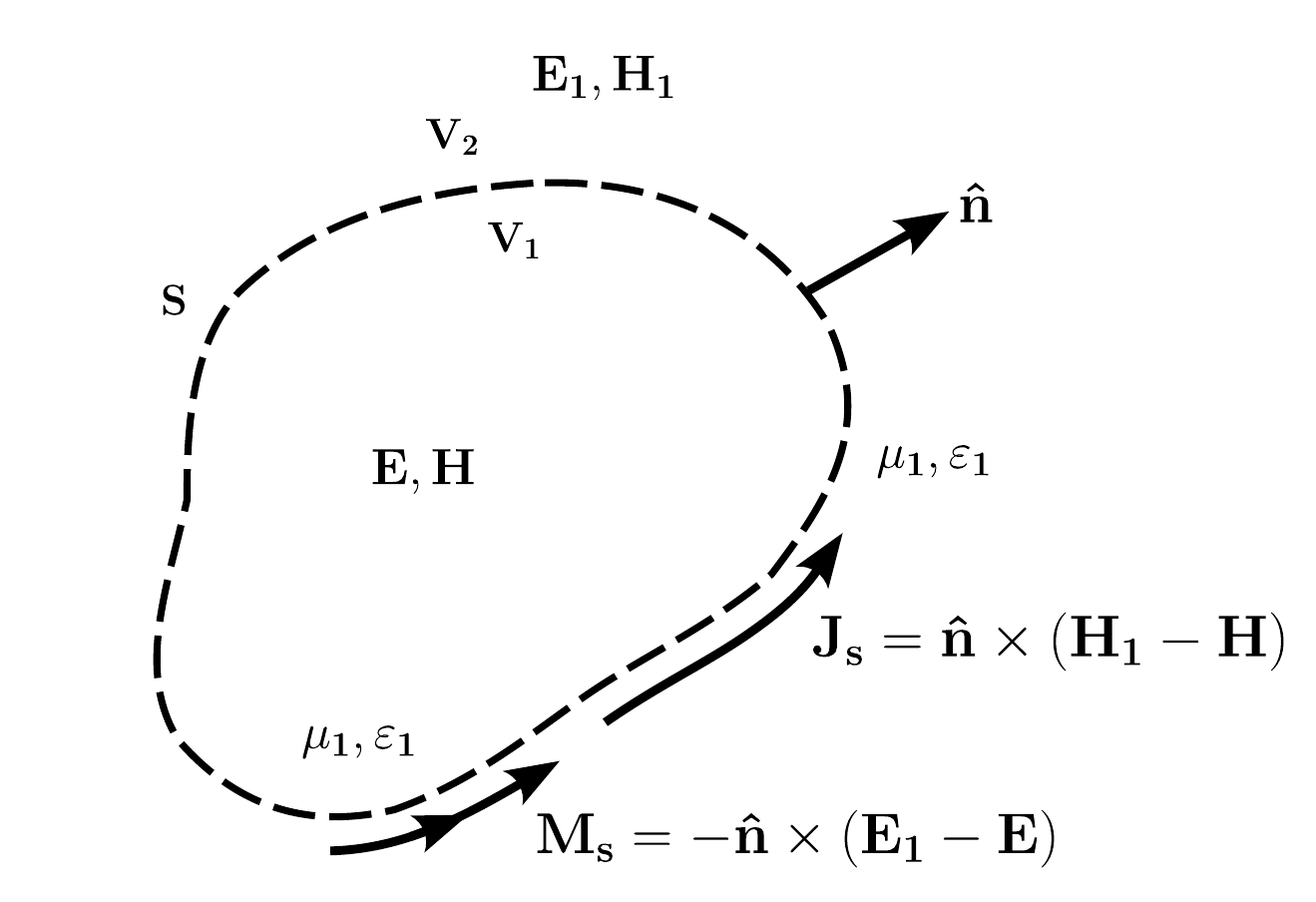} }}%
    \caption{(a) Actual and (b) equivalent surface model \cite{Balanis2016}.}%
    \label{fig:equivalence_model}%
\end{figure}

The Surface Equivalence Theorem considers an actual radiating source represented by its electric $\mathbf{J}_s$  and magnetic $\mathbf{M}_s$ current densities as shown in Fig.\ref{fig:equivalence_model}.
The sources radiate fields $\mathbf{E}_1$ and $\mathbf{H}_1$ everywhere. For these fields to exist within and outside of the surface $S$, they must satisfy the boundary conditions on the tangential electric and magnetic field components as shown in Fig. \ref{fig:equivalence_model}(b)
\begin{equation}  -\hat{\mathbf{n}} \times {( \mathbf{E}_1 - \mathbf{E}) = \mathbf{M}_s} \label{eq1}\end{equation}
\begin{equation}  \hat{\mathbf{n}} \times {( \mathbf{H}_1 - \mathbf{H}) = \mathbf{J}_s}            \label{eq2}\end{equation}
Since the fields inside of the surface $S$ is not the region of interest, it can be assumed that they are zero. Therefore (\ref{eq1}) and (\ref{eq2}) reduce to
\begin{equation}  \mathbf{M}_s = -\hat{\mathbf{n}} \times { \mathbf{E}_1} \label{eq3}\end{equation}
\begin{equation}   \mathbf{J}_s = \hat{\mathbf{n}} \times { \mathbf{H}_1}            \label{eq4}\end{equation}
This form of the field equivalence principle is known as Love's equivalence principle \cite{AEEBalanis2024}.

A microstrip patch antenna can be modelled as a dielectric-loaded cavity with two perfect conducting electric walls representing the patch and ground plane, and four perfectly conducting magnetic walls representing the side apertures of the antenna. From the surface equivalence theorem the patch antenna can be represented by equivalent electric and magnetic current densities.

\begin{figure} [ht]
    \centering
    \subfloat[\centering ]{{\includegraphics[width=8cm, trim={0cm 0cm 0cm 0cm},clip]{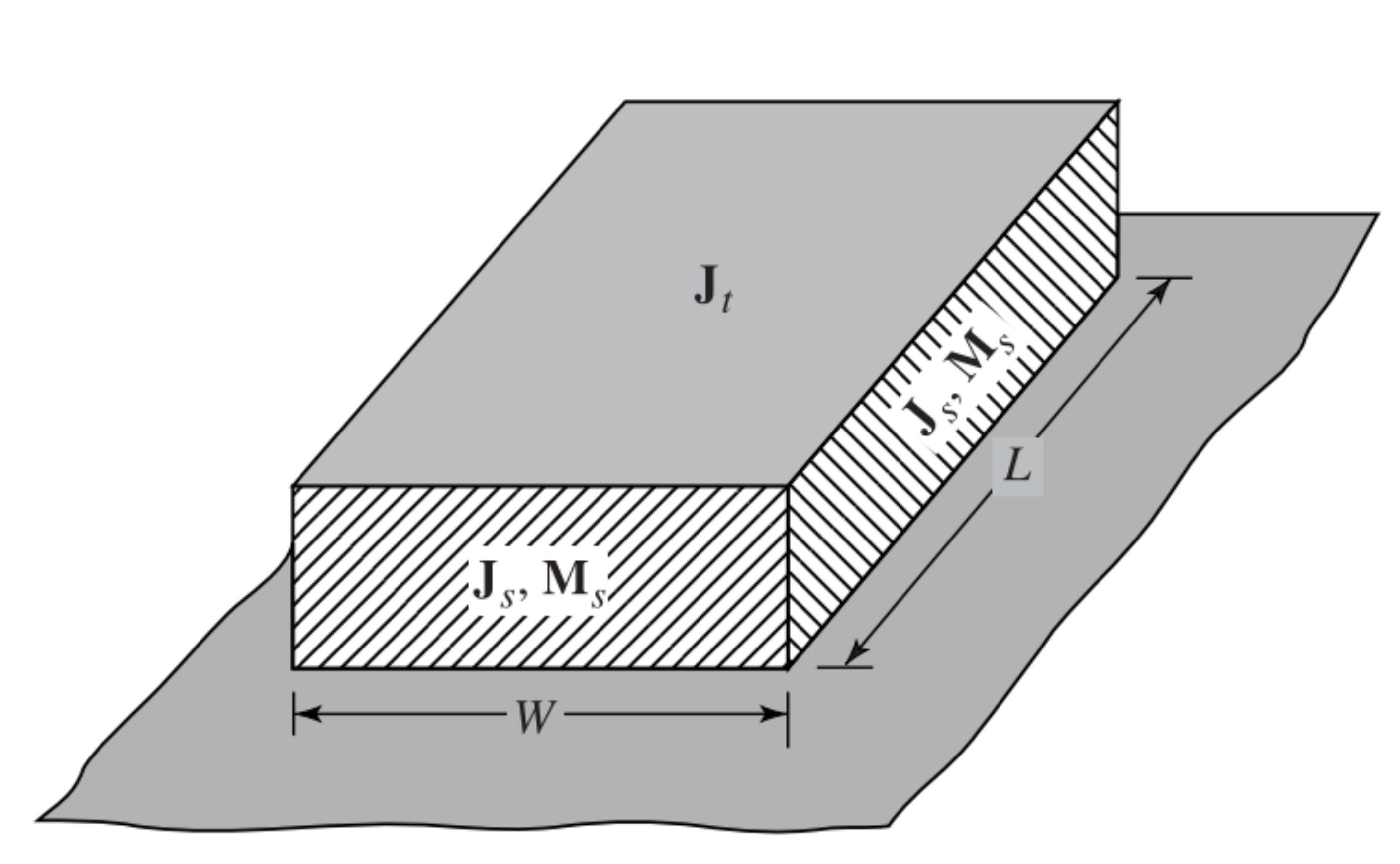} }}%
    \vspace{0.01cm}
    \subfloat[\centering ]{{\includegraphics[width=6.5cm, trim={0cm 0cm 0cm 0cm},clip]{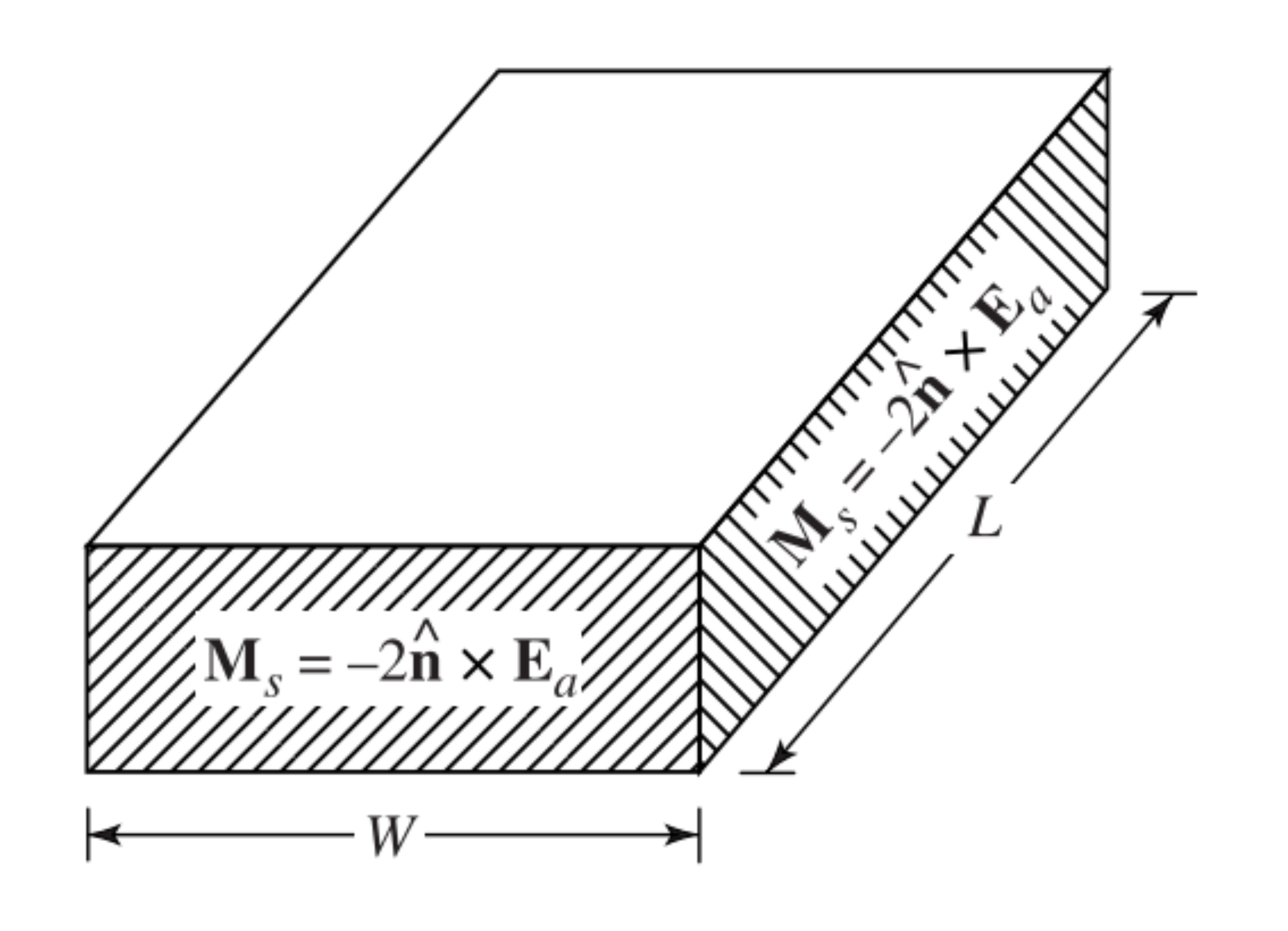} }}%
    \caption{Patch antenna (a) current densities and (b) equivalent current densities model \cite{Balanis2016}.}\label{patchequivs}
\end{figure} 

As shown in Fig.\ref{patchequivs}(a) the top and bottom of the patch antenna can be represented by the current density $\mathbf{J}_t$, while the side apertures are represented by an electric current density $\mathbf{J}_s$ and magnetic current density $\mathbf{M}_s$. In practice, patch antennas are designed where the height to width ratio is small which results in a small current density $\mathbf{J}_t$ \cite{Balanis2016} and will be assumed to be zero in this analysis. Since the tangential magnetic fields along the edges of the patch are small $\mathbf{J}_s$ is also assumed to be zero. The only nonzero current density is the magnetic current densities $\mathbf{M}_s$ and is doubled to account for the antenna's ground plane by use of the image theory. C. Balanis \cite{Balanis2016} showed that this model is a good approximation characterising the normalised electric and magnetic field distributions.

\section{Proposed Cavity Model For A Patch Antenna Embedded With A PMC} \label{proposed cavity model}

The use of the Cavity Model for farfield radiation characterisation dates back to the late 1970s and early 1980s \cite{Lo1979} \cite{Carver1981}. For the proposed RPA-HGP, embedding a PMC layer halfway between the patch antenna's radiating edges produces an equivalent electric and magnetic current density across its apertures. Therefore, the total fields propagated from the aperture is modified compared to that of a conventional patch antenna. In this section the analytical foundations to form a farfield analysis are outlined for the antenna model.
To find the magnetic and electric fields generated by either an electric current source $\mathbf{J}$ and a magnetic current $\mathbf{M}$, the inhomogeneous vector wave equations in the form of the magnetic and electric auxiliary potential functions $\mathbf{A}$ and $\mathbf{F}$ are required to be solved, which are given by
\begin{equation} \nabla^2 \mathbf{A} + k^2\mathbf{A} = -\mu \mathbf{J} \label{magnetic_aux}\end{equation}
\begin{equation} \nabla^2 \mathbf{F} + k^2\mathbf{F} = -\varepsilon \mathbf{M} \label{electric_aux}\end{equation}
where $k$ is the wavenumber, $\mathbf{\mu}$ is the permeability and $\mathbf{\varepsilon}$ is the permittivity of the aperture. The process to derive the solution to these equations is found in the Appendix.
\begin{figure} [!ht]
    \centering
    \subfloat[\centering ]{{\includegraphics[width=8cm, trim={0cm 0cm 0cm 0cm},clip]{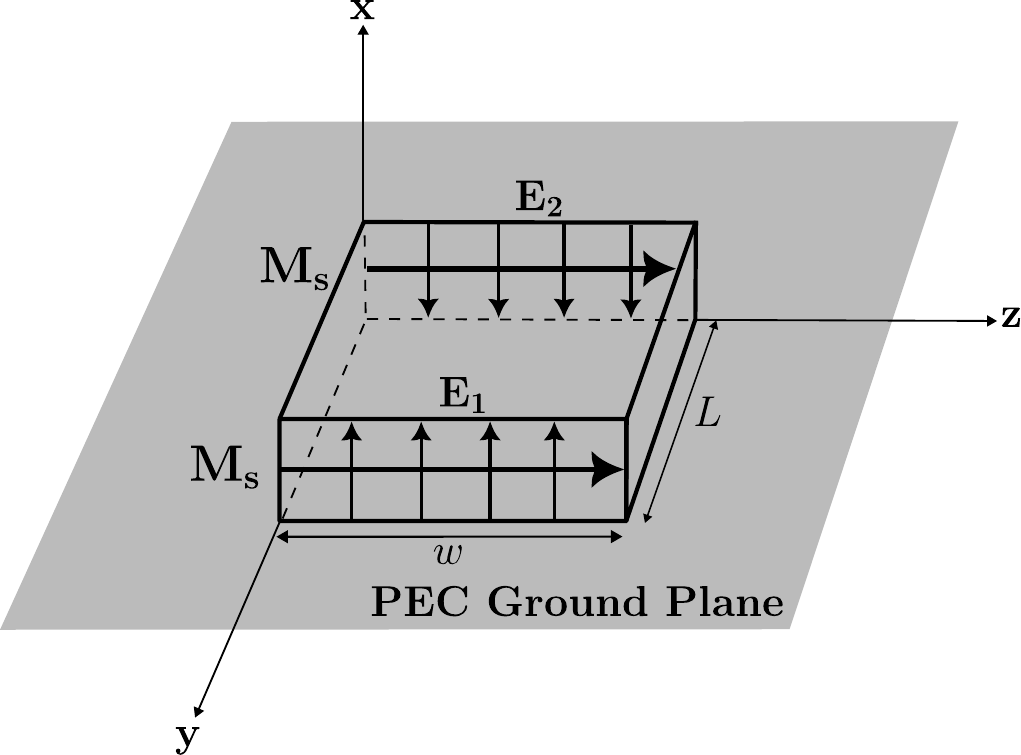} }}%
    \vspace{0.01cm}
    \subfloat[\centering ]{{\includegraphics[width=8cm, trim={0cm 0cm 0cm 0cm},clip]{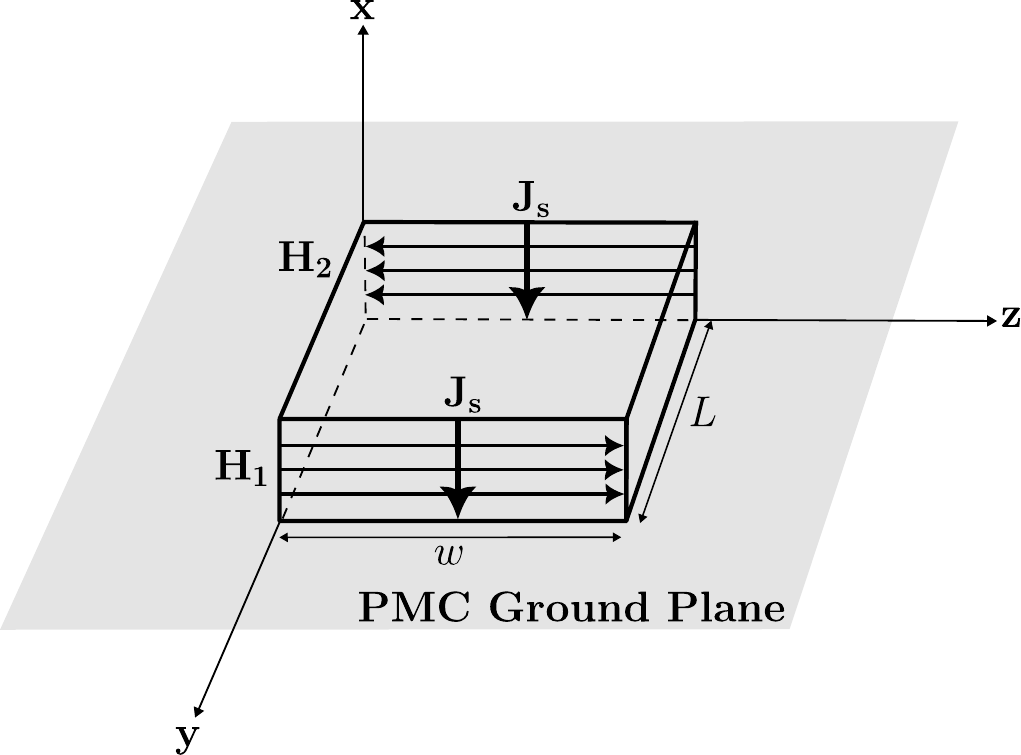} }}%
    \caption{Patch antenna geometry when placed on a (a) PEC and (b) PMC ground plane.}
    \label{fig:patch_geometry_pec_pmc}%
\end{figure} 
The proposed optimised rectangular patch antenna (RPA) is represented by four radiating apertures or slots. However, two of the side apertures cancel along the principal planes. These apertures are known as the non-radiating edges and thus are not included in the analysis. The main radiating apertures have an embedded PMC layer in the middle of the aperture while the bottom of the aperture lays on a PEC ground plane. To simplify the problem the aperture placed on the PEC and PMC ground plane are considered individually as shown in Fig.\ref{fig:patch_geometry_pec_pmc}, and then later combined as the superposition of both the induced current densities. This means that the electric $E_a$ and magnetic $H_a$ fields propagated from the main radiating apertures will be due to an induced magnetic $\mathbf{M_s}$ and electric $\mathbf{J_s}$ surface current density, which take the form
\begin{equation}
    \mathbf{M}_s = - 2\hat{\mathbf{n}} \times E_a
\end{equation}
\begin{equation}
    \mathbf{J}_s = 2\hat{\mathbf{n}} \times H_a
\end{equation}


The current densities are doubled due to the presence of the PMC and PEC ground layers by the Image Theory \cite{AEEBalanis2024}.
Based on the geometry from Fig.\ref{fig:patch_geometry_pec_pmc}(a) the electric field across the aperture is modelled as a uniform distribution, therefore 
\begin{equation}
  E_a =
    \begin{cases}
      \hat{\mathbf{a}}_z E_0 & \text{$-h \leq x \leq h$ , $-w/2 \leq z \leq w/2$ }\\
      0 & \text{elsewhere}
    \end{cases}       
\end{equation}
From Fig.\ref{fig:patch_geometry_pec_pmc} the induced current densities are given as 
\begin{equation}
    \mathbf{M}_s = - 2\hat{\mathbf{n}} \times E_a = -\hat{\mathbf{a}}_y \times  \hat{\mathbf{a}}_x 2E_0 = +\hat{\mathbf{a}}_z 2E_0
\end{equation}
\begin{equation}
    \mathbf{J}_s = 2\hat{\mathbf{n}} \times H_a =  -\hat{\mathbf{a}}_y \times\left(\hat{\mathbf{a}}_z\frac{2E_0}{\eta_a}\right) = -\hat{\mathbf{a}}_x \frac{2E_0}{\eta_a}
\end{equation}
where $\eta_a$ is the wave impedance at the aperture, relating the electric field to the magnetic field amplitude.

\begin{figure}[!hb] 
    \centering
    \includegraphics[width=\linewidth, trim={0 0 0 0},clip]{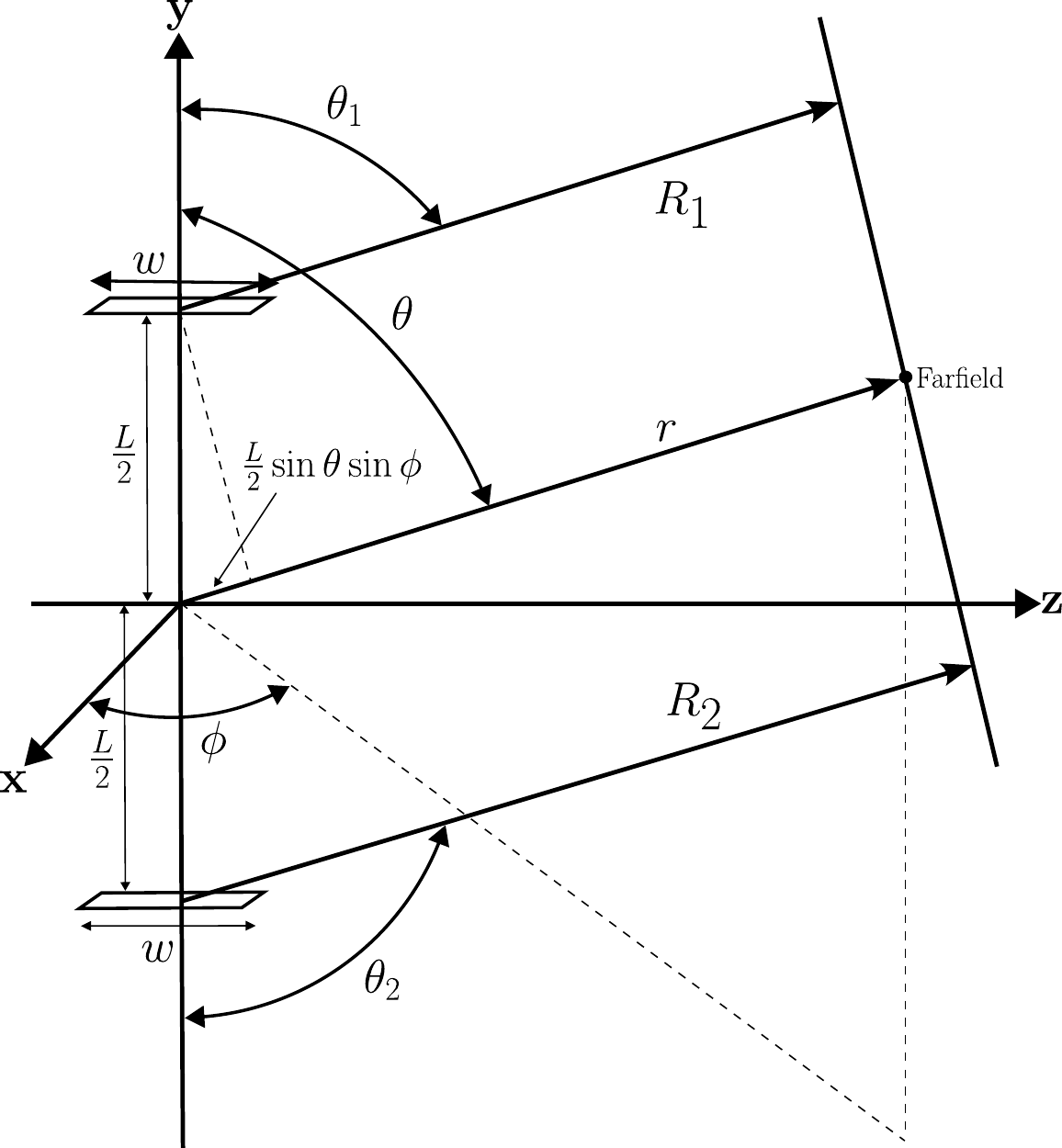}
    \caption{Aperture array geometry for farfield approximation.}
    \label{fig:farfield_geometry}
\end{figure}

In order to derive the equations for the electric field components, the space factor terms $\mathbf{N}$ and $\mathbf{L}$ are computed as given by (\ref{electric_sf}) and (\ref{magnetic_sf}) in the Appendix. The differential path $r'\cos{\psi}$ in the space factor equations is calculated from the difference in the paths from the source to the observation point as
\begin{equation}
    r'\cos{\psi} = \mathbf{r}' \cdot \hat{\mathbf{a}}_r = x'\sin{\theta}\cos{\phi} + z'\cos{\theta}
    \label{diff_path}
\end{equation} The electric field along the $\theta$ component $E_\theta$ is found from (\ref{E_theta}) in the Appendix by computing  $L_\phi$ and $N_\theta$. Since there is only a magnetic current density along the $z$ component, $L_\phi = 0$. Since the PMC layer is centred in the middle of the aperture, the height of its image is half of the overall aperture height, therefore the integral limits for computing $N_\theta$ is from $\-h/2< x' <h/2$. By substituting the differential path given by (\ref{diff_path}) into (\ref{N_theta}), the space factor $N_\theta$ is given as
\begin{multline}
   N_\theta = \cos{\theta}\cos{\phi} J_x \int_{-\frac{h}{2}}^{\frac{h}{2}} e^{jk(x'\sin{\theta}\cos{\phi})} \,dx' \\ \int_{-\frac{w}{2}}^{\frac{w}{2}} e^{jk(z'\cos{\theta})} \,dz'
\end{multline}
Computing the integral gives us the space factor $N_\theta$ in the form
\begin{equation}
    N_\theta = \cos{\theta}\cos{\phi} J_x \sinc \left(
    {X}\right) \sinc \left({Z}\right)
\end{equation}
where $X = {\frac{kh}{2} \sin{\theta}\cos{\phi} } $ and $Z = {\frac{kw}{2} \cos{\theta}}$. Since the aperture's height is much smaller than the operating wavelength of the antenna, the $\sinc({X})$ variations are small and therefore are neglected in this analysis. Therefore, $E_\theta$ is expressed as
\begin{equation}
    E_\theta = j\frac{khwE_0}{2\pi}\frac{\eta}{\eta_a}\cos{\theta}\cos{\phi}\sinc{(Z)} \frac{e^{-jkr}}{r}
    \label{E_theta_final}
\end{equation}
$E_\phi$ is found from (\ref{E_phi}) by computing the space factors $L_\theta$ and $N_\phi$. The image of the aperture without the PMC layer has a height $h$, therefore when computing $L_\theta$ the limits of integration for the height component is from $\-h<x'<h$. However, the height variations are small and are assumed negligible.
By repeating the same procedure that was used to derive (\ref{E_theta_final}),  $E_\phi$ becomes
\begin{equation}
    E_\phi = j\frac{khwE_0}{2\pi} \left(\frac{\eta}{\eta_a}\sin{\phi} - \sin{\theta}\right) \sinc{(Z)} \frac{e^{-jkr}}{r}
    \label{E_phi_final}
\end{equation}
Fig.\ref{fig:farfield_geometry} outlines the geometry of the aperture array. The radial distance $R_1$ and $R_2$ are defined as the distance from each aperture to the farfield observation point. Both apertures are symmetrically positioned from the origin with a length of $\pm L/2$ along the $y$-axis.

The radial distance $R_1$ from the aperture positioned at $+L/2$ is calculated as
\begin{equation}
    R_1 = \sqrt{x^2 + \left(y-\frac{L}{2}\right)^2 + z^2}
\end{equation}
Using the binomial expansion on the $y$ term to the first 2 terms, substituting $r^2 = x^2 + y^2 + z^2$ and $y=r\cos{\psi}$, the radial distance can be simplified to
\begin{equation}
    R_1 \approx r - \frac{L}{2}\cos{\psi}
    \label{R1}
\end{equation}
For the aperture positioned at $-L/2$, the radial distance $R_2$ is calculated as
\begin{equation}
    R_2 \approx r + \frac{L}{2}\cos{\psi}
    \label{R2}
\end{equation}
where $\psi$ is the angle between the axis of the aperture and the radial vector from the origin to the farfield observation point. Since the aperture elements are placed along the $y$ axis, $\cos{\psi}$ is transformed to $\sin{\theta}\sin{\phi}$.
To account for the radiation produced by each aperture at the farfield observation point, equations (\ref{R1}) and (\ref{R2}) are substituted into the phase term $e^{-jkr}$ of (\ref{E_theta_final}) and are added together since the total field is the superposition of the two apertures. Therefore, the phase term becomes
\begin{equation}
\begin{split}
     e^{jkr} &= e^{-jkR_1} + e^{-jkR_2} \\
            & = e^{-jk\left( r - \frac{L}{2}\sin{\theta}\sin{\phi} \right)} + e^{-jk\left( r + \frac{L}{2}\sin{\theta}\sin{\phi} \right)} \\
            & = 2\cos\left({\frac{kL}{2}\sin{\theta}\sin{\phi}}\right) e^{jkr}  
            \label{phase_term}
\end{split}
\end{equation}
By implementing (\ref{phase_term}) to (\ref{E_theta}) and (\ref{E_phi}), the final form of $E_\theta$ and $E_\phi$ are given as
\begin{equation}
    E_\theta = j\frac{khwE_0}{2\pi}\frac{\eta}{\eta_a}\cos{\theta}\cos{\phi}\sinc{(Z)} \left(AF_y \right) \frac{e^{-jkr}}{r}
    \label{E_theta_final_AF}
\end{equation}
\begin{equation}
    E_\phi = j\frac{khwE_0}{2\pi} \left(\frac{\eta}{\eta_a}\sin{\phi} - \sin{\theta}\right) \sinc{(Z)} \left(AF_y \right)\frac{e^{-jkr}}{r}
    \label{E_phi_final_AF}
\end{equation}
where $\left(AF_y \right)$ is known as the array factor and is given by
\begin{equation}
    \left(AF_y \right) = 2\cos\left({\frac{kL}{2}\sin{\theta}\sin{\phi}}\right)
\end{equation}
To simplify the analysis, the wave impedance in free space $\eta$ and across the aperture $\eta_a$ relating the magnetic field to the electric field are equal.
The total electric field $\mathbf{E}$ is expressed as
\begin{equation}
    \mathbf{E} = \hat{\mathbf{n}}_r E_r + \hat{\mathbf{n}}_\theta E_\theta + \hat{\mathbf{n}}_\phi E_\phi 
    \label{total_e_field}
\end{equation}
The magnitude $\mathbf{\left|E \right|}$ can be written as
\begin{equation}
    \mathbf{\left|E \right|} = \sqrt{ \left |E_\theta \right|^2 + \left |E_\phi \right|^2}
    \label{magnitude}
\end{equation}
$E_r\approx0$ since its amplitude varies inversely proportional to $r^2$, and therefore at the farfield distance becomes much smaller than $E_\theta$ and $E_\phi$.
The normalised electric field magnitude $\mathbf{E_n}$ is given as
\begin{equation}
    \mathbf{E_n} = \frac{\mathbf{\left|E\left(\theta ,\phi \right)\right|}}{\left |\mathbf{E}_{max} \right|}
    \label{norm_magnitude}
\end{equation}
In order to compute the normalised electric field pattern, an approximation for the width $w$ of the apertures and length $L$ between the apertures is required. For the farfield computation, the aperture's length is an effective length, which means that there's a length extension added to account for the fringing of the fields. Therefore the length $L$ in the array factor $\left(AF_y \right)$ is replaced with the effective length $L_{eff}$. 

This analysis is compared to a conventional rectangular patch antenna whereby only a magnetic current density is induced over the aperture. By performing the same analysis outlined above it can be shown that $E_\theta = 0$ and $E_\phi$ is given as

\begin{equation}
      E_\phi = j\frac{khwE_0}{2\pi} \sin{\theta}\sinc{(Z)} \left(AF_y \right) \frac{e^{-jkr}}{r}
\end{equation}
The value for the width of the conventional patch is approximated using 
\begin{equation} \label{patch_width}
    w = \frac{c}{2f_r}\sqrt{\frac{2}{\varepsilon_r +1}}
\end{equation}
as this equation is shown to produce good radiation efficiencies \cite{bahl1980microstrip}, where $f_r$ is the resonant frequency of the antenna and $c$ is the speed of light.
The length is approximated using \cite{Balanis2016}
\begin{equation} \label{analytical_length}
    L = \frac{1}{2f_r \sqrt{\varepsilon_{reff}}\sqrt{\mu_0 \varepsilon_0}} - 2\Delta L
\end{equation}
where $\varepsilon_{reff}$ is the effective dielectric constant which accounts for the fringing fields produced by the patch apertures  and is given by \cite{Balanis2016}
\begin{equation}
    \varepsilon_{reff} = \frac{\varepsilon_r + 1}{2} + \frac{\varepsilon_r - 1}{2}\left[1+12\frac{h}{w} \right]^{-\frac{1}{2}}
\end{equation}
The incremental length $\Delta L$  corresponds to the length extension that occurs due to the fringing fields produced at the apertures, which makes the antenna look electrically larger than its length $L$. $\Delta L$ is approximated by \cite{bahl1980microstrip}
\begin{equation}
    \Delta L = 0.412 h \frac{\left(\varepsilon_{reff} + 0.3 \right)\left( \frac{w}{h}+0.264\right)}{\left(\varepsilon_{reff} - 0.258 \right)\left( \frac{w}{h}+0.8\right)}
\end{equation}
Thus, the effective lengh $L_{eff}$ is given as
\begin{equation}
    L_{eff} = L + 2\Delta L
\end{equation}
Currently there is no analytical method to predict the length and width for the RPA-HGP. Therefore, approximations need to be made for the dimensions that best fit the simulated results. This is discussed in section \ref{amc_integration}.
\section{Designing the AMC cell array} \label{designing AMC array}
\begin{figure}[ht] 
    \centering
    \includegraphics[width=\linewidth, trim={0 0cm 0 0cm},clip]{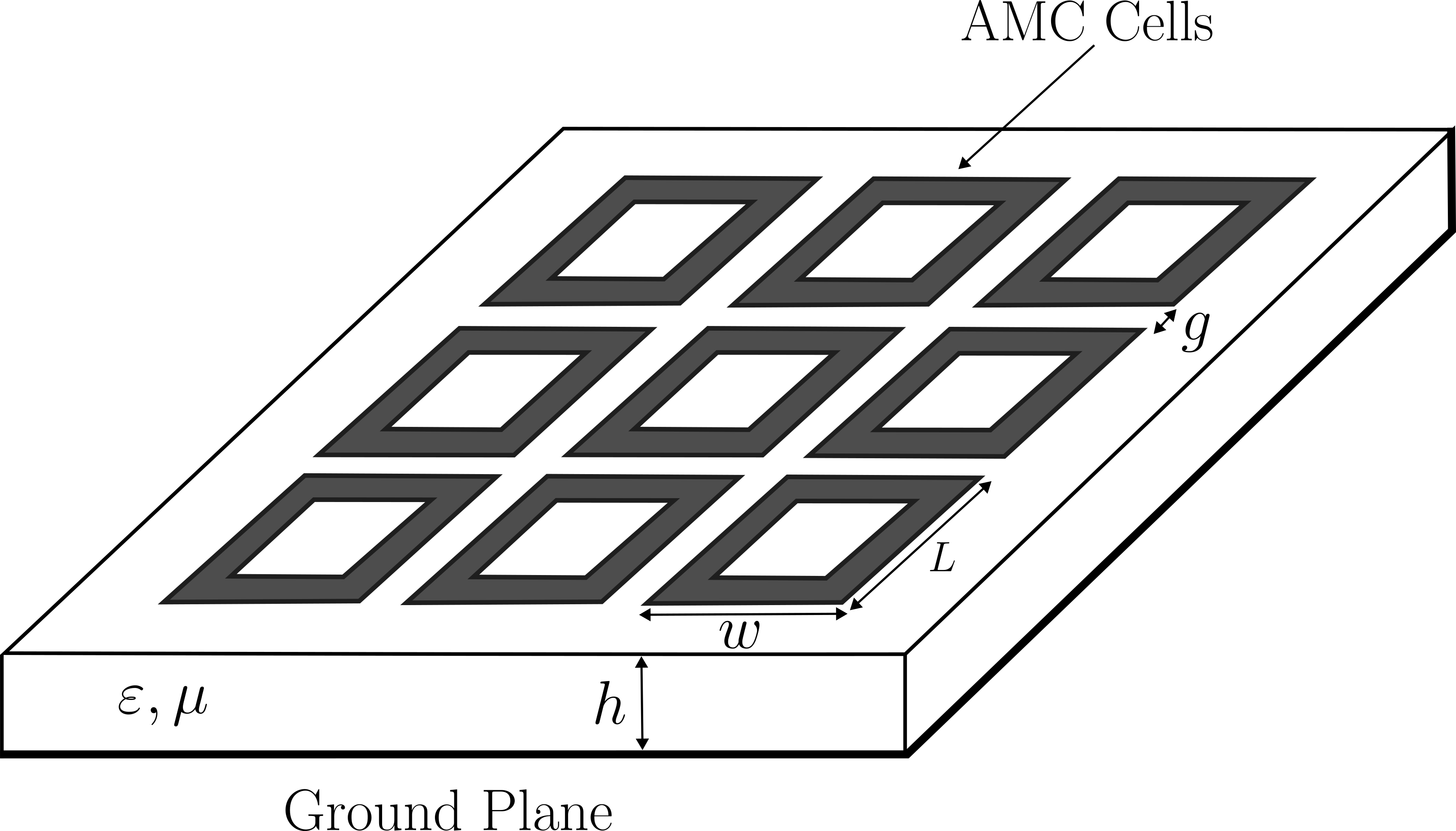}
    \caption{Geometry of the AMC surface.}
    \label{fig:amc_array}
\end{figure}

AMC cells need to be designed to exhibit the PMC properties within the apertures of the patch antenna. So far the analytical model presented assumes an infinite PMC ground layer. To realise the PMC ground plane properties a geometry and array spacing of the individual scatterers needs to be considered. Sievenpiper et al. \cite{Sievenpiper2000} had analysed a hexagonal and square patch geometry of metal plates for the HIS. 
In this research, a square loop is used as the AMC surface, since its resonance is at approximately $\lambda/4$. This results in a smaller overall profile of the unit cells when combined in an array configuration as well as a lower profile when integrated into the patch antennas substrate. The geometry of the proposed surface is given in Fig.\ref{fig:amc_array}. 

\begin{figure}[ht] 
    \centering
    \includegraphics[width=7.5cm, trim={0 0cm 0 0cm},clip]{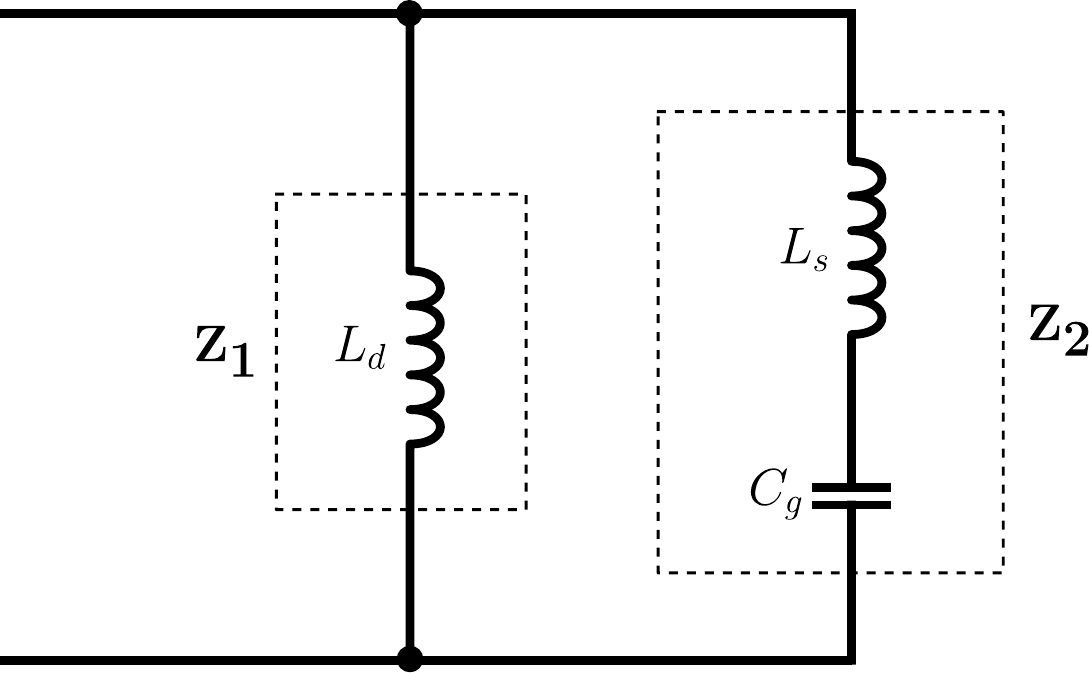}
    \caption{AMC unit cell circuit equivalent model.}
    \label{fig:equiv_circuit}%
\end{figure}

\begin{figure} [ht]
    \centering
    \begin{tikzpicture}
        \begin{axis}[
        xmin=1.6,xmax=3.4,
        xtick={1.6,1.8,2,2.2,2.4,2.6,2.8,3,3.2,3.4},
        ytick={-180,-135,-90,-45,0,45,90,135,180},
        grid = both,
        xlabel={ $\mathrm{Frequency\,(GHz)}$}, 
        ylabel={$\mathrm{Reflection\;Phase\,(dB)}$},
        ylabel style={yshift=-4pt},
        width = 8.2cm,
        legend entries = {Equivalent Circuit, Simulation},
        legend pos = south west,
        legend style={nodes={scale=0.65}}]
    
    \addplot[black,dashed,very thick] table{./data/amc_phase_equiv_model.txt}; \label{RPA}
    \addplot[black,very thick] table{./data/amc_phase_simulation.txt}; \label{RPA-AMC}
        \end{axis}
    \end{tikzpicture}
    \caption{Circuit equivalent and full-wave simulation phase reflection profile of the square AMC surface.}
    \label{fig:phase_reflect}

\end{figure}
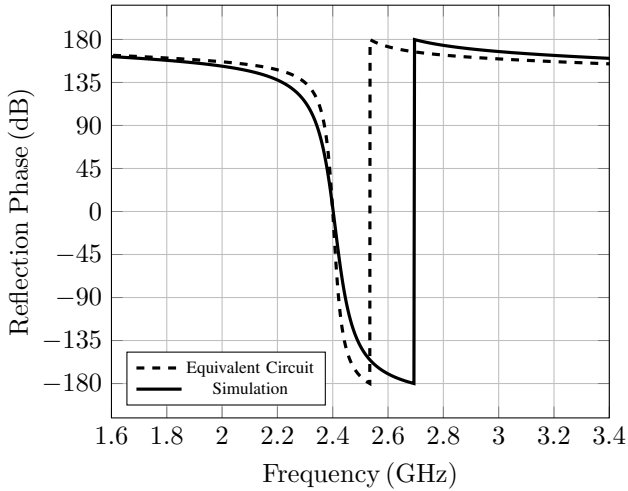

The parameter of interest for the AMC surface is its phase reflection. In order to analytically characterise this for the square loop array, a surface equivalent circuit model is created and analysed. Consider an incident TM (or TE) wave on a unit cell of the AMC array as shown in Fig.\ref{fig:amc_array} characterised by a surface impedance $Z_s$, the reflection coefficient $R$ is given as
\begin{equation}
    R = \frac{Z_s - \eta}{Z_s + \eta}
\end{equation}
where $\eta$ is the free space impedance. When the wave is incident on an electrical wall (PEC), the surface impedance $\left|Z_s \right| = 0$, therefore the reflection coefficient is $R=-1$, which means that the reflected wave has a phase which is reversed when compared to the incident wave. For a magnetic wall (PMC), the surface impedance is infinite $\left|Z_s \right| = \infty$ and thus the reflection coefficient becomes $R=+1$, which implies the reflected wave has the same phase as the incident wave. While the AMC cannot achieve a infinite surface impedance, their absolute values are large enough to exhibit PMC properties.
To illustrate the phase reflection properties of the proposed AMC surface, a circuit equivalent model is created as shown in Fig.\ref{fig:equiv_circuit}. The circuit model consists of a sheet inductance $L_s$ due to the square loop AMC geometry and a sheet capacitance $C_g$ due to the gap between the adjacent square loop. The sheet inductance and capacitance are in series together. Since the thickness of the layer between the array of the AMC cells and the ground plane is much smaller than the operating wavelength, the input impedance on the thin layer is inductive \cite{Tretyakov2003a}. This small inductance $L_d$ is attributed by the ground plane backing the unit cell, and is parallel to $L_s$ and $C_g$.
The inductance $L_d$ is calculated as
\begin{equation}
    L_d = \mu h
\end{equation}
where $h$ is the height of the substrate and $\mu$ is the mediums permeability.
The total surface impedance $Z_s$ is calculated as a parallel impedance between the square ring AMC array and the ground plane which is given as 
\begin{equation}
    Z_s^{-1} = Z_1^{-1} + Z_2^{-1}
\end{equation}
where $Z_1 = j\omega L_d$ and $Z_2 = j\omega L_s + \frac{1}{j\omega C_g}$, the total surface impedance $Z_s$ can be simplified and expressed as
\begin{equation}
    Z_s = \frac{j\omega L_d \left(L_s C_g \omega^2 -1 \right)}{C_g\omega^2 \left(L_d + L_s \right)-1}
\end{equation}
The resonant frequency $\omega_0$ of the equivalent circuit is given by summing the reactive components and equating it to zero. It is expressed as
\begin{equation}
    \omega_0 = \frac{1}{\sqrt{C_g\left(L_d + L_s \right)}}
    \label{loop_fres}
\end{equation}
The inductance $L_s$ of the square loop is calculated as \cite{Paul2010}
\begin{equation}
    L_s = \frac{2\mu_0L}{\pi}\left[\ln\left({\frac{L}{r_w}}\right)-0.774\right]
\end{equation}
where $L$ is the length of the square loop and $r_w$ is the radius of the loop. Since the square loop is assumed to be two dimensional being placed on a flat surface with a very small height, $r_w$ is approximated as half the thickness of the loop.
The gap capacitance between each unit cell can be approximated by rearranging (\ref{loop_fres}) to
\begin{equation}
    C_g = \frac{1}{\omega^2 \left(L_d + L_s \right)}
\end{equation}
Finally, the reflection phase $\varphi$ from the circuit model of the AMC surface is calculated as
\begin{equation}
    \varphi = \operatorname{Im}\left\{\ln{\left(\frac{Z_s-\eta}{Z_s + \eta} \right)} \right\} 
\end{equation}
The length of the square loop is approximated with a length of $\lambda/4$. The designer chooses the operating frequency of the AMC cells, the square loop's thickness, the height of the separation between the AMC cells and the ground plane and finally the gap separation between the AMC cells. The square loop's thickness is arbitrarily chosen as $5$mm wide which means $r_w = 2.5$mm. To simplify the analysis the substrate is modelled as free space. Fig.\ref{fig:phase_reflect} shows the phase reflection profile of the equivalent circuit model. The operating bandwidth of the AMC surface is over the regions where the phase reflection is between $-90\degree$ and $+90\degree$. Therefore, the bandwidth of the sqaure ring AMC array using the circuit equivalent model is approximately $2.1\%$ or $50$MHz. 

\subsection{Full-Wave Simulation Comparison} \label{full-wave simulation}
The square ring AMC cell is designed in CST Microwave Studio. CST allows unit cell analysis using a ‘Floquet port’ which allows planes of periodicity modelling by exciting linearly polarized TM and TE plane waves. The boundary condition defined around the edges of the structures is called a ‘Unit Cell’ boundary condition, which repeats the modelled structure periodically in two directions up to infinity. The incident fields across the unit cell is the same at every point. A perfect electrical conductor boundary condition is defined at the backside of the AMC unit cell which acts as a ground plane.
For a direct comparison to the equivalent circuit model, the metallic components of the AMC unit cell have material definition of a PEC. The surrounding substrate is modelled as free space, thus $\varepsilon_r\approx1$. The AMC square loop is modelled as a thin sheet which means that it is a 2-dimensional structure with negligible height. The simulated dimensions are given as $L=w=31.7$mm, $h=4$mm, $g=0.6$mm. The simulated bandwidth is $4.2\%$ or $100$MHz. 

\section{Integrating the AMC surface into the patch antenna} \label{amc_integration}
The proposed configuration is to place the AMC layer in the middle of the substrate just like that of the PMC layer in the hybrid ground plane for the Cavity Model presented earlier. The dimensions are based on the simulated unit cell analysis presented in section \ref{designing AMC array}. The antenna is designed to operate at $2.4$GHz and is directly compared to a conventional RPA, which is designed using the transmission line model outlined by Balanis \cite{Balanis2016}. 
\begin{figure} []
    \centering
    \subfloat[\centering ]{{\includegraphics[width=8.7cm, trim={0cm 0.5cm 0cm 0cm},clip]{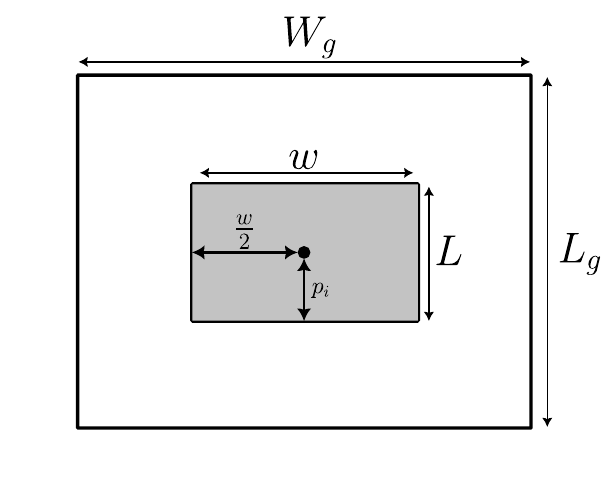} }}%
    \vspace{0cm}
    \subfloat[\centering ]{{\includegraphics[width=8.7cm, trim={0cm 0.4cm 0cm 0cm},clip]{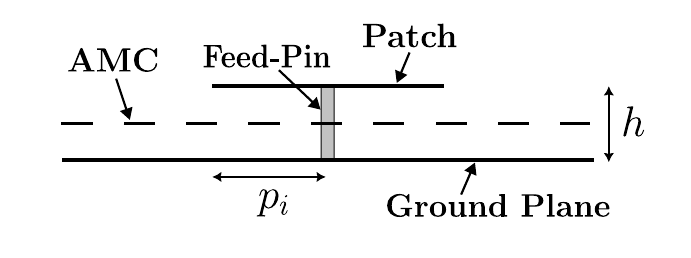} }}%
    \caption{(a) Front view and (b) side view patch dimensions showing the embedded AMC surface. The conventional RPA has a width $w = 62.4$mm, length $L = 55.7$mm and a feed-pin inset distance $p_i = 16$mm. The RPA-AMC has a width $w = 40$mm, a length $L = 37.9$mm and a feed-pin inset distance $p_i = 1$mm. The height and ground plane dimensions of each antenna design are given as $h = 4$mm, $L_g = W_g = 160$mm.}
    \label{fig:patch_views}%
\end{figure} 
The conventional RPA is designed and simulated in CST. The parameters used to design the dimensions are optimised to produce a resonance of $2.4$GHz. The dimensions for both antennas are outlined in Fig.\ref{fig:patch_views}. The antennas are fed using a centre fed coaxial cable through a hole in the ground plane and antenna substrate to the radiating element. The feed point is initially placed in the centre of the radiating element and its location is varied across the y-axis by altering the value of the inset $p_i$ to control the input impedance.

The patch antenna with the embedded AMC surface (RPA-AMC) is initially designed with the patch dimensions of the conventional RPA. It is then optimised to exhibit a resonance at $2.4$GHz. In order to realise the hybrid ground plane layer modelled in the Cavity Model which is assumed to be an infinitely large structure, a large 4$\times$5 AMC cell array is designed on top of a ground plane with dimensions $L_g = W_g = 160$mm = $1.12\lambda$, with the patch element designed at the centre. The simulated S-parameters of the proposed RPA-AMC is shown in Fig.\ref{fig:s_parameters_comp}. A comparison is made to a conventional RPA designed and simulated to operate at $2.4$GHz. It is shown that the RPA-AMC has a resonance at $2.4$GHz with patch dimensions of $0.30\lambda\times 0.32\lambda$, while the conventional RPA has dimensions of $0.45\lambda\times 0.5\lambda$ showing that there is a miniaturisation of the antenna's length by $28.2\%$ and width by $27.9\%$. This is attributed due to the AMC layer exhibiting a large permittivity at resonance. The bandwidth for the RPA is $3.3\%$ $(2.35-2.43\:\mathrm{GHz})$ while the bandwidth for the RPA-AMC is $2.1\%$ $(2.36-2.41 \:\mathrm{GHz})$, which shows an approximate $30.5\:\mathrm{MHz}$ reduction in bandwidth for the RPA-AMC when compared to a conventional RPA. Further bandwidth may be gained by optimising the feed-pin location to obtain a better impedance match. The simulated S-parameter shows resonances between $2.6-3\:\mathrm{GHz}$ which are coupling resonances between the feedline and the AMC array.

\begin{figure} [h!]
    \centering
    \begin{tikzpicture}

    \begin{axis} 
    [xmin=2,xmax=3, ymax = 0,
    grid = both, 
    xlabel={ $\mathrm{Frequency\,(GHz)}$}, 
    ylabel={$\mathrm{Return\;Loss\,(dB)}$},
    ylabel style={yshift=-6pt},
    width = 8cm,
    legend entries = {RPA,RPA-AMC}, 
    legend pos = south east,
    legend style={nodes={scale=0.7}}]

    \addplot[black,dashed,very thick] table{./data/rpa_s11.txt}; \label{RPA}
    \addplot[black,very thick] table{./data/rpa_amc_s11.txt}; \label{RPA-AMC}
    \end{axis}
    \end{tikzpicture}
    \caption{Simulated S-parameter comparison between the RPA and RPA-AMC}
    \label{fig:s_parameters_comp}
\end{figure}
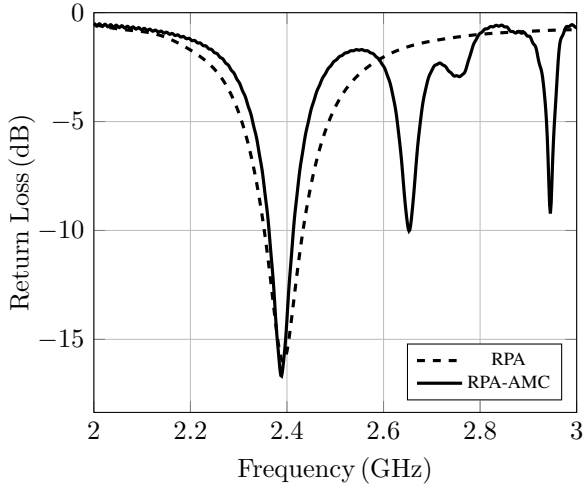

\begin{figure} [h!]
    \centering
    \begin{tikzpicture}
    \begin{axis}
    [xmin=0,xmax=90, ymax = 0, ymin = -40,
    xtick={0,10,20,30,40,50,60,70,80,90},
    grid = both, 
    xlabel={ $\mathrm{\phi(\degree)}$}, 
    ylabel={$\mathrm{|{E(\theta,\phi)} / {E_{max}} |\; {(dB)} }$},
    ylabel style={yshift=-6pt},
    width = 8cm,
    legend entries = {Cavity Model RPA,Cavity Model RPA-HGP, Simulated RPA,Simulated RPA-AMC}, 
    legend pos = south west,
    legend style={nodes={scale=0.7}}]
    
    \addplot[black, dashed,very thick] table{./data/cavity_model_rpa_eplane.txt};
    \label{Cavity Model RPA}
    \addplot[black, very thick] table{./data/cavity_model_rpa_amc_eplane.txt};
    \label{Cavity Model RPA-HGP}
    \addplot[black, densely dashdotdotted, very thick] table{./data/sim_rpa_eplane.txt};
    \label{Simulated RPA}
    \addplot[black, dotted, very thick] table{./data/sim_rpa_amc_eplane.txt};
    \label{Simulated RPA-AMC}

    \end{axis}
    \end{tikzpicture}

    \caption{E-plane comparison between Cavity Model and simulation}
    \label{fig:e_plane_comp}
\end{figure}
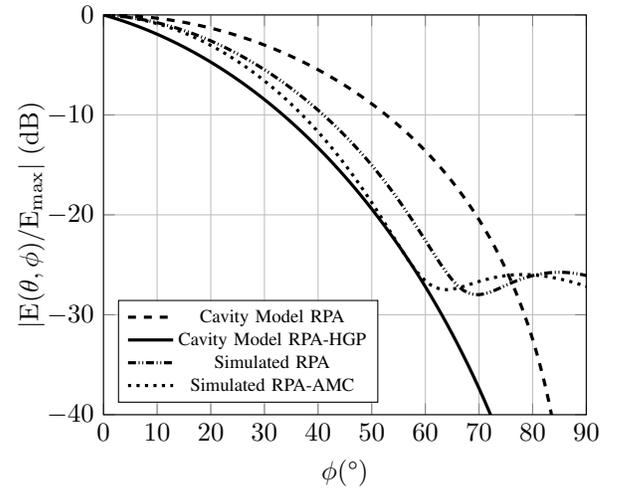

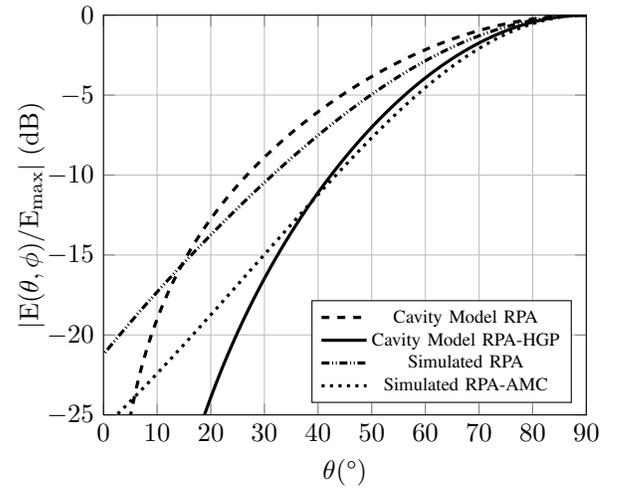
\begin{figure} [ht]
    \centering
    \begin{tikzpicture}
    \begin{axis} 
    [xmin=0,xmax=90, ymax = 0, ymin = -25,
     xtick={0,10,20,30,40,50,60,70,80,90},
    grid = both, 
    xlabel={ $\mathrm{\theta(\degree)}$}, 
    ylabel={$\mathrm{|{E(\theta,\phi)} / {E_{max}} |\; {(dB)} }$},
    ylabel style={yshift=-6pt},
    width = 8cm,
    legend entries = {Cavity Model RPA,Cavity Model RPA-HGP, Simulated RPA,Simulated RPA-AMC}, 
    legend pos = south east,
    legend style={nodes={scale=0.7}}]
    
    \addplot[black, dashed,very thick] table{./data/cavity_model_rpa_hplane.txt};\label{Cavity Model RPA}
    \addplot[black,very thick] table{./data/cavity_model_rpa_amc_hplane.txt};\label{Cavity Model RPA-HGP}
    \addplot[black, densely dashdotdotted, very thick] table{./data/sim_rpa_hplane.txt};
    \label{Simulated RPA}
    \addplot[black, dotted, very thick] table{./data/sim_rpa_amc_hplane.txt};
    \label{Simulated RPA-AMC}

    \end{axis} 
    \end{tikzpicture}

    \caption{H-plane comparison between Cavity Model and simulation.}
    \label{fig:h_plane_comp}
\end{figure}

\begin{figure} []
    \centering
    \subfloat[\centering ]{{\includegraphics[width=8cm, trim={0cm 0cm 0cm 0cm},clip]{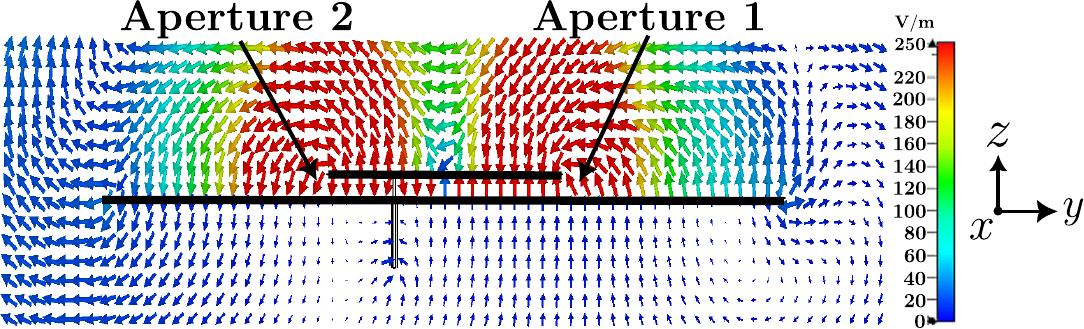} }}%
    \vspace{0.5cm}
    \subfloat[\centering ]{{\includegraphics[width=8cm, trim={0cm 0cm 0cm 0cm},clip]{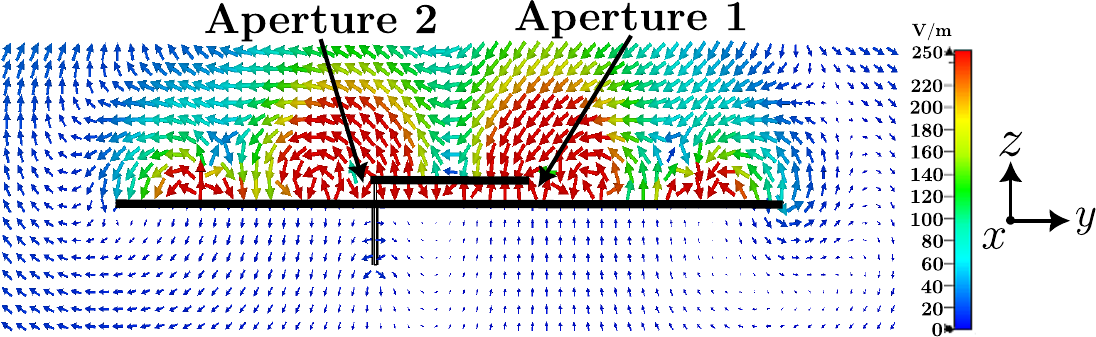} }}%
    \caption{E-field distribution at $2.4$GHz across the  E-plane of the (a) RPA and (b) the RPA-AMC with an excitation phase $\psi=90\degree$.}
    \label{fig:e_field_cuts}%
\end{figure}

\begin{figure} [!ht]
    \centering
    \subfloat[\centering ]{{\includegraphics[width=8cm, trim={0cm 0cm 0cm 0cm},clip]{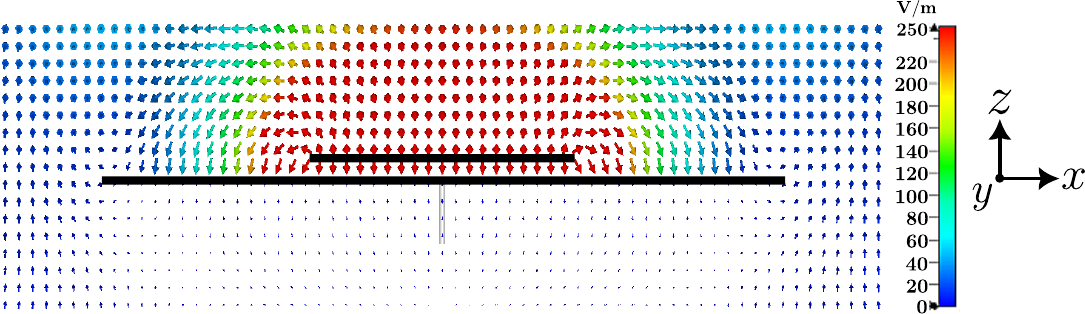} }}%
    \vspace{0.5cm}
    \subfloat[\centering ]{{\includegraphics[width=8cm, trim={0cm 0cm 0cm 0cm},clip]{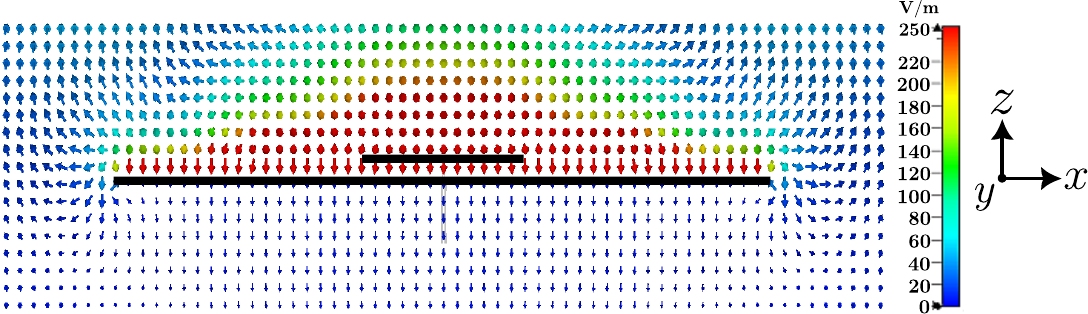} }}%
    \caption{E-field distribution at $2.4$GHz across the H-plane of the (a) RPA and (b) the RPA-AMC with an excitation phase $\psi=90\degree$. The H-plane cut is centred across Aperture 2 at the point where the maximum E-field distribution is located.}
    \label{fig:h_field_cuts}%
\end{figure}

The simulated farfield plots are shown in Fig.\ref{fig:e_plane_comp}-\ref{fig:h_plane_comp} and are compared to the analytical analysis using the cavity model. The plots are represented as a normalised electric field pattern for a direct comparison. For the cavity model, the antenna's length is approximated using (\ref{analytical_length}), while the antenna's width is adjusted to best fit the simulated result since (\ref{patch_width}) is shown to be not valid for this antenna. This is due to the addition of the AMC array which couples strongly around the antenna's apertures. This results in creating an electrically larger aperture width. For this antenna it is shown that an effective aperture width of approximately $\lambda$ results in a good match. The effective widths and lengths are illustrated in the E-field plots given in Fig. \ref{fig:e_field_cuts}-\ref{fig:h_field_cuts}. 

For the RPA-AMC, the farfield patterns predict that RPA-AMC will be more directive compared to a conventional patch. This is shown to be in agreement with the simulated E-plane pattern. However, using the simulated width of the patch for the analytical model shows a predicted H-plane pattern which is very broad and not a good match to the simulated result which shows that the H-plane is more directive. It appears that the effective width spans the width of the AMC surface plus a slight width extension due to the coupling of the AMC cells. The simulated directivity of the RPA is $10.5\:\mathrm{dBi}$ while the RPA-AMC's directivity is $11.9\:\mathrm{dBi}$ which is a $1.4\:\mathrm{dBi}$ increase when compared to the RPA, which agrees with the predicted result using the Cavity Model. At approximately $\phi=60\degree$ and $\phi=90\degree$ in the E-plane plot, the RPA-AMC and RPA farfield pattern starts to diverge when compared to the Cavity model. This is due to edge diffraction effects that occur for a finite ground plane. The divergence can also be seen in the H-plane but to a lesser extent.

\section{Conclusion}
The Cavity Model is presented for a patch antenna embedded with a hybrid ground in order to predict its farfield characteristics. The PMC layer within the hybrid ground plane is realised through a full wave simulation model using an AMC surface composed of square rings. A circuit equivalent model is developed to model its phase reflection properties. The Cavity Model predicts an increase in the patch antenna's directivity. This is portrayed by plotting the normalised E-field farfield pattern, whereby a narrower distribution implies a higher directivity. The Cavity Model is shown to be in agreement with the simulated results. Additionally, through the full-wave simulation results, the RPA-AMC shows a miniaturisation of the radiating element's length and width of approximately $28\%$.

\appendix[Method to solve the auxiliary inhomogenous vector wave equations]
The details for constructing the solution to solve the auxiliary inhomogenous vector wave equations are adapted and referenced from Harrington \cite{Harrington2001} and Balanis \cite{AEEBalanis2024}.

The magnetic auxiliary potential function $\mathbf{A}$ for an electric source $\mathbf{J}$ is related to the electric $\mathbf{E}$ and magnetic fields $\mathbf{H}$ by
\begin{equation} \mathbf{E} \approx -j\omega \mathbf{A} \label{E_field_aux}\end{equation}
\begin{equation} \mathbf{H} \approx -j \frac{\omega}{\mu}  \hat{\mathbf{a}}_r \times \mathbf{A} \label{H_field_aux}\end{equation}
For a magnetic source $\mathbf{M}$ they are related to the electric $\mathbf{E}$ and magnetic fields $\mathbf{H}$ by
\begin{equation} \mathbf{H} \approx -j\omega \mathbf{F} \label{E_field_aux_2}\end{equation}
\begin{equation} \mathbf{E} \approx -j \omega\mu  \hat{\mathbf{a}}_r \times \mathbf{F} \label{H_field_aux_2}\end{equation}
In this model, the sources are represented as linear densities, therefore the solutions to the wave equation can reduce to surface integrals as
\begin{equation}
    \mathbf{A} = \frac{\mu}{4\pi} \iint_s \mathbf{J} \frac{e^{-jkR}}{R} \,ds' 
\end{equation}
\begin{equation}
    \mathbf{F} = \frac{\varepsilon}{4\pi} \iint_s \mathbf{M} \frac{e^{-jkR}}{R} \,ds' 
\end{equation}
Assuming the current density resides on the surface of the source, $\mathbf{A}$ and  $\mathbf{F} $ will take the form
\begin{equation}
    \mathbf{A} = \frac{\mu}{4\pi} \iint_s \mathbf{J} \frac{e^{-jkR}}{R} \,ds' \approx \frac{\mu e^{j k r}}{4\pi r} \mathbf{N} 
\end{equation}

\begin{equation}
    \mathbf{F} = \frac{\varepsilon}{4\pi} \iint_s \mathbf{M} \frac{e^{-jkR}}{R} \,ds' \approx \frac{\varepsilon e^{j k r}}{4\pi r} \mathbf{L} 
\end{equation}
where $\mathbf{N}$ and $\mathbf{L}$ are known as the space factors, which determines the electric and magnetic current densities along the source observed at the farfield, given as
\begin{equation}
    \mathbf{N} =  \iint_s \mathbf{J_s} e^{jkr'\cos{\psi}} ds'
    \label{electric_sf}
\end{equation}
\begin{equation}
    \mathbf{L} =  \iint_s \mathbf{M_s} e^{jkr'\cos{\psi}} ds'
    \label{magnetic_sf}
\end{equation}
Expanding equations (\ref{electric_sf}) and (\ref{magnetic_sf}) in a rectangular coordinate system yields
\begin{equation}
    \mathbf{N} =  \iint_s (\mathbf{ \hat{\mathbf{a}}}_x J_x + \hat{\mathbf{a}}_y J_y + \hat{\mathbf{a}}_z J_z ) e^{jkr'\cos{\psi}} ds'
    \label{electric_sf_expanded}
\end{equation}
\begin{equation}
    \mathbf{L} =  \iint_s (\mathbf{ \hat{\mathbf{a}}}_x M_x + \hat{\mathbf{a}}_y M_y + \hat{\mathbf{a}}_z M_z ) e^{jkr'\cos{\psi}} ds'
    \label{magnetic_sf_expanded}
\end{equation}
The product of the element factor and the space factor gives the total radiated field. The element factor refers to the term outside of the brackets in (\ref{E_theta}), (\ref{E_phi}), (\ref{H_theta}) and (\ref{H_phi}) along with any term that can be factored out of the integral for 

For the farfield analysis $L_\phi$, $L_\theta$, $N_\theta$ and $N_\phi$ are transformed to a spherical co-ordinate system using the transformation matrix
\begin{equation}
   \begin{bmatrix}
\hat{\mathbf{a}}_x \\
\hat{\mathbf{a}}_y \\
\hat{\mathbf{a}}_z
\end{bmatrix}
= \begin{bmatrix}
    \sin{\theta}\cos{\phi} & \cos{\theta}\cos{\phi} & -\sin{\phi}\\
    \sin{\theta}\sin{\phi} & \cos{\theta}\sin{\phi} & \cos{\phi} \\
    \cos{\theta} & -\sin{\theta} & 0
    \end{bmatrix} 
\begin{bmatrix}
\hat{\mathbf{a}}_r \\
\hat{\mathbf{a}}_\theta \\
\hat{\mathbf{a}}_\phi
\end{bmatrix}
\end{equation}
Doing so yields
\begin{equation} 
    N_\theta = \iint_s  \begin{bmatrix}
    \cos{\theta}\cos{\phi} \\ 
    \cos{\theta}\sin{\phi} \\
    -\sin{\theta}    
    \end{bmatrix}
    \begin{bmatrix}
        J_x \\ J_y \\ J_z
    \end{bmatrix}
    e^{jkr'\cos{\psi}} ds'
    \label{N_theta}
\end{equation}
\begin{equation} 
    N_\phi = \iint_s  \begin{bmatrix}
    -\sin{\phi} \\ 
    \cos{\phi} \\
    0    
    \end{bmatrix}
    \begin{bmatrix}
        J_x \\ J_y \\ J_z
    \end{bmatrix}
    e^{jkr'\cos{\psi}} ds'
     \label{N_phi}
\end{equation}
\begin{equation} 
    L_\theta = \iint_s  \begin{bmatrix}
    \cos{\theta}\cos{\phi} \\ 
    \cos{\theta}\sin{\phi} \\
    -\sin{\theta}    
    \end{bmatrix}
    \begin{bmatrix}
        M_x \\ M_y \\ M_z
    \end{bmatrix}
    e^{jkr'\cos{\psi}} ds'
     \label{L_theta}
\end{equation}
\begin{equation} 
    L_\phi = \iint_s  \begin{bmatrix}
    -\sin{\phi} \\ 
    \cos{\phi} \\
    0    
    \end{bmatrix}
    \begin{bmatrix}
        M_x \\ M_y \\ M_z
    \end{bmatrix}
    e^{jkr'\cos{\psi}} ds'
    \label{L_phi}
\end{equation}

In the farfield region only the $\theta$ and $\phi$ components are included while the radial component is negligible due to being very small \cite{AEEBalanis2024}.
The electric fields in terms of their auxiliary potentials can be expressed as
\begin{equation}
    (E_A)_\theta \approx -j\omega A_\theta 
\end{equation}
\begin{equation}
    (E_A)_\phi \approx -j\omega A_\phi 
\end{equation}
\begin{equation}
    (E_F)_\theta \approx -j\omega \eta F_\phi 
\end{equation}
\begin{equation}
    (E_F)_\phi \approx -j\omega \eta F_\theta 
\end{equation}
and for the magnetic fields
\begin{equation}
    (H_F)_\theta \approx -j\omega F_\theta 
\end{equation}
\begin{equation}
    (H_F)_\phi \approx -j\omega F_\phi 
\end{equation}
\begin{equation}
    (H_A)_\theta \approx j\omega \frac{A_\phi}{\eta}  
\end{equation}
\begin{equation}
    (H_A)_\phi \approx -j\omega \frac{A_\theta}{\eta}   
\end{equation}
where $\eta$ is the free space wave impedance.
Thus, the total electric $\mathbf{E}$ and magnetic $\mathbf{H}$ fields can be written as the superposition between the electric and magnetic auxiliary potential fields due to an electric $\mathbf{J}$ and magnetic $\mathbf{M}$ current density source given as
\begin{equation}
    E_r \approx 0 
\end{equation}
\begin{equation}
    E_\theta \approx -\frac{jke^{-jkr}}{4\pi r}(L_\phi + \eta N_\theta)
    \label{E_theta}
\end{equation}
\begin{equation}
    E_\phi \approx \frac{jke^{-jkr}}{4\pi r}(L_\theta - \eta N_\phi)
    \label{E_phi}
\end{equation}
\begin{equation}
    H_r \approx 0 
\end{equation}
\begin{equation}
    H_\theta \approx \frac{jke^{-jkr}}{4\pi r}(N_\phi - \frac{L_\theta}{\eta} )
    \label{H_theta}
\end{equation}
\begin{equation}
    H_\phi \approx -\frac{jke^{-jkr}}{4\pi r}(N_\theta + \frac{L_\phi}{\eta} )
    \label{H_phi}
\end{equation}




\ifCLASSOPTIONcaptionsoff
  \newpage
\fi

\bibliographystyle{IEEEtran}
\bibliography{ref.bib}
\end{document}